# A two-phase argon avalanche detector operated in a single electron counting mode

A. Bondar[a], A. Buzulutskov[*,a], A. Grebenuk[a],
D. Pavlyuchenko[a], R. Snopkov[a], Y. Tikhonov[a],
V.A. Kudryavtsev[b], P.K. Lightfoot[b], N.J.C. Spooner[b]

[a] *Budker Institute of Nuclear Physics, 630090 Novosibirsk, Russia*

[b] *Dept. of Physics and Astronomy, University of Sheffield, Hicks Building, Hounsfield Road, Sheffield, S3 7RH, UK*

**Abstract**

The performance of a two-phase Ar avalanche detector in a single electron counting mode was studied, with regard to potential application in coherent neutrino-nucleus scattering and dark matter search experiments. The detector comprised of a 1 cm thick liquid Ar layer and a triple-GEM multiplier operated in the saturated vapour above the liquid phase. Successful operation of the detector in single electron counting mode, in the gain range from 6000 to 40000, has for the first time been demonstrated.

## 1. Introduction

The development of detectors having a low energy threshold and low noise is of paramount importance for present-day and future low-background experiments. In particular, in experiments searching for dark matter [1] and coherent neutrino-nucleus scattering [2,3], the detector should be sensitive to rather weak ionization and scintillation signals produced by nuclear recoils. Cryogenic two-phase detectors in general [1] and specifically two-phase Ar detectors [3,4,5] are currently favoured for these applications due to their capability of efficient background rejection.

The weak scintillation and ionization signals in such two-phase detectors need to be amplified. This can be achieved using either PMTs [1] or Gas Electron Multipliers (GEMs, [6]) [7]. In the latter case the detector is operated in an electron avalanching mode and is called a Two-Phase Avalanche Detector. Its successful performance is provided by the unique ability of multi-GEM structures to operate in all noble gases at high gains at room [8,9] and cryogenic temperatures

[*] Corresponding author. Email: buzulu@inp.nsk.su

[7,10,11,12,13], including in the two-phase mode in Ar, Kr and Xe [7,10,11,12].

In Ar, the average nuclear recoil energy is below 1 keV and 100 keV for coherent neutrino [2,3] and dark matter [4,5] induced scattering, respectively. Since only a fraction of the recoil energy converts into ionization and excitation (of the order of 20% at low energies [3]), the primary ionization signal may consist of only a few electrons. This is particularly relevant for coherent neutrino scattering, where the primary ionization signal would typically be a single electron signal [3]. Thus, in this type of experiment the cryogenic avalanche detector must operate in a single electron counting mode, which is a real challenge.

It has been recently demonstrated that a two-phase Ar avalanche detector with a triple-GEM multiplier can be operated at rather high gains (in particular compared to two-phase Kr and Xe avalanche detectors), reaching $10^4$ [12]. Optimizing electronic noise appropriately, the magnitude of this gain might be high enough for the efficient detection of single electrons.

In this paper, the performance of a two-phase Ar avalanche detector in a single electron counting mode is studied for the first time. We show that a triple-GEM multiplier in two-phase Ar can effectively detect single electrons in the gain range from several thousands to several tens of thousands.

## 2. Experimental setup and procedures

A detailed description of the experimental setup has been presented elsewhere [10,11,12]. Here we describe details relevant to the detector performance in a single electron counting mode. A cathode mesh, three GEM foils and a printed circuit board (PCB), of an active area of $28\times28$ mm$^2$ each, were mounted in a cryogenic vacuum-insulated chamber of volume 2.5 l (figure 1). The distance between the cathode and the first GEM was 12 mm, between the GEMs - 2 mm and between the third GEM and the PCB - 2 mm. In two-phase mode the detector was operated close to the triple point, at a temperature of 84 K and a vapour pressure of 0.70 atm. At this point, the thickness of the liquid condensate at the chamber bottom was approximately 1.1 cm, which is a factor of 3 larger than that in the previous measurements [12]. The electron drift path in liquid Ar before attachment was found to be about 1 cm at a drift field of 2 kV/cm.

The detector was operated in a pulse-counting mode using a pulsed X-ray tube. The anode voltage on the tube was 50 kV. The width and frequency of X-ray pulses were 400 ns (FWHM) and 250 Hz respectively.

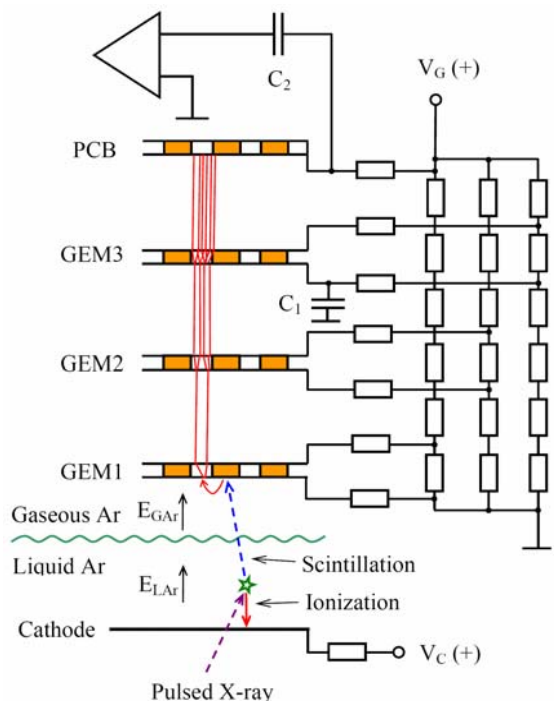

Figure 1. Schematic view of the experimental setup to study the performance of the two-phase Ar avalanche detector with the triple-GEM multiplier in single electron counting mode (not to scale). The 3GEM+PCB configuration is depicted. The drift field is reversed compared to the standard configuration.

The triple-GEM multiplier was operated in either a 3GEM or 3GEM+PCB configuration. In the 3GEM+PCB configuration, the anode signal was read out from the PCB, as shown in figure 1. In the 3GEM configuration, the anode signal was recorded from the last electrode of the third GEM. In this case the PCB was under floating potential and the field lines were terminated at the last electrode of the third GEM, providing for full collection of the avalanching electrons. In contrast, in the 3GEM+PCB configuration only 1/3 of the electrons are collected at the PCB. On the other hand, the 3GEM+PCB configuration is more practical for position-sensitive readout. Therefore we studied both readout configurations.

The cathode, GEM and PCB electrodes were biased through a resistive high-voltage divider placed outside the cryostat (figure 1). Each electrode was connected to the divider using high-voltage feedthroughs and 1 m long wires.

The anode signal was read out from the divider using a charge-sensitive preamplifier followed by a research amplifier with an overall sensitivity of 11 V/pC and shaping time of 10 µs. In the current work, all the measurements were done in a triggered mode, the trigger signal being provided by the X-ray generator.

The reduction of electronic noises, pick-ups and HV source ripples is of primary importance for the operation in single electron counting mode. Therefore the following measures were taken to reduce them: all voltages were supplied to the divider through second-order RC filters; the line voltages were powered through ferrite-ring inductance filters; the HV divider and the preamplifier were placed in a shielded box; the external wires used for signals prior to amplification were shielded; the ground positions were optimized.

Let us call the electric field in the cathode gap in the liquid ($E_{LAr}$) and in the gas above the liquid ($E_{GAr}$) as the drift field in general (see figure 1). Note, that the drift fields in the liquid and in the gas phase are different:

$$E_{LAr} = V_C / (d_L + d_G \varepsilon_L / \varepsilon_G)$$
$$E_{GAr} = V_C / (d_L \varepsilon_G / \varepsilon_L + d_G) \quad (1).$$

Here $V_C$ is the voltage applied to the cathode gap, $d_L$ and $d_G$ the liquid and gas layer thickness, $\varepsilon_L$ and $\varepsilon_G$ the liquid and gas dielectric constants.

The standard direction of the drift field is defined as that at which the electrons drift from the cathode towards the first GEM, i.e. the same as that in the GEM holes and in the transfer and induction gaps. The standard field direction was used when measuring the gain characteristics and operation stability of the detector. In single electron counting mode, however, the drift field was reversed, in order to suppress primary ionization signals and to have only scintillation-induced photoelectric signals producing single electrons at the first GEM (see figure 1 and section 4 for details).

## 3. Gain and stability characteristics

The gain values used in the following were determined with the standard direction of the drift field: the gain is defined as the pulse height of the avalanche (anode) signal divided by that of the calibration signal. The calibration signal was recorded at the first electrode of the first GEM, with no high voltage applied to the divider.

Fig.2 shows the measured gain characteristic of the triple-GEM in two-phase mode at 84 K and 0.70 atm. For comparison, the gain characteristic at room temperature at 1.9 atm, in the gaseous mode, is also shown. Both characteristics are similar to each other in terms of the slopes, confirming the statement that the avalanche mechanisms in the gas and two-phase modes are essentially the same.

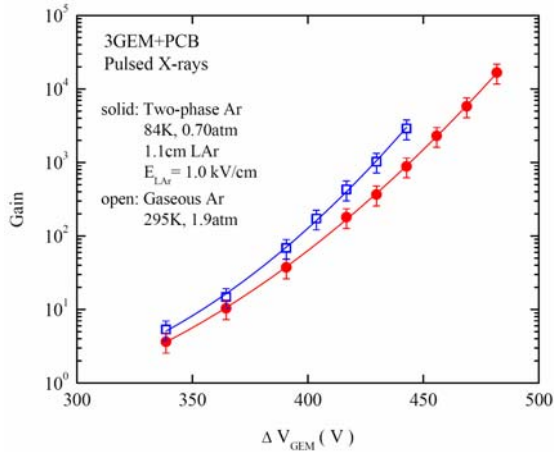

Figure 2. Gain of the triple-GEM in the 3GEM+PCB configuration as a function of the voltage across each GEM, in two-phase Ar at 84 K and 0.70 atm and in gaseous Ar at 295 K and 1.9 atm. The drift field has the standard direction and is equal to $E_{LAr}$=1.0 kV/cm. The maximum gains are not reached.

In two-phase mode rather high gains were obtained in the 3GEM+PCB configuration, of about $2\times10^4$. In the 3GEM configuration, even higher gains were obtained, reaching about $4\times10^4$. These gains are somewhat higher than those reported earlier [12] and certainly high enough for successful performance in a single electron counting mode.

The operation stability of two-phase avalanche detectors is a principal question. For instance, it has been reported that when using a wire chamber [14] or Micromegas readout [15], the signal from the two-phase Xe avalanche detector disappears within a few tens of minutes. On the other hand, stable operation of a GEM-based two-phase avalanche detector was observed for several hours in Ar [12] and Kr [11] and for half an hour in Xe [12], with the liquid layer thickness of 3 mm.

Here we confirm stable operation of a two-phase Ar detector, having a thicker liquid layer than that in the previous measurements. Figure 3 illustrates the operational stability of the detector over a period of 5 hours: the pulse-height from the triple-GEM at a gain of $5\times10^2$ is shown as a function of time when irradiated with pulsed X-rays at regular intervals. The measurements were started just after the detector reached the pressure stabilization point during the cooling cycle [12]. This point is close to the triple point of Ar.

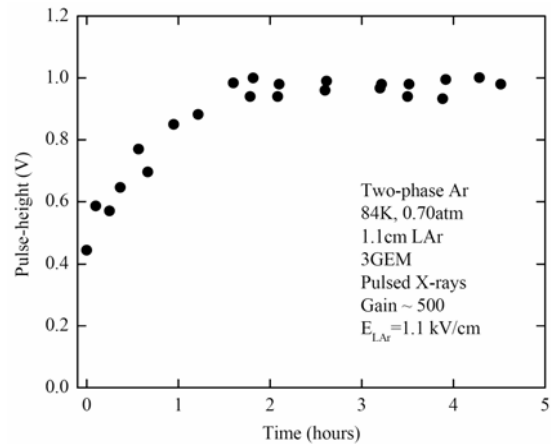

Figure 3. Stability of operation of the two-phase Ar avalanche detector at 84 K and 0.70 atm. Anode pulse-height from the triple-GEM (3GEM) at a gain of $5\times10^2$ is shown as a function of the time. The signals are induced by pulsed X-rays. The drift field has the standard direction and is equal to $E_{LAr}$=1.1 kV/cm.

One can see that the pulse-height is stabilized after two hours. The initial stabilization period is most probably due to the time needed to smooth the temperature gradients in the cryostat. Taking this into account, one may conclude that the operation of the two-phase Ar avalanche detector is rather stable.

## 4. Single electron counting mode

The standard practice for studying single electron counting mode is to use a light source, with the intensity gradually decreased down to a single photon level, the single electrons being produced at the photocathode coupled to the detector. In our case, however, a mere decrease of the intensity of the X-ray tube does not result in single electron counting since each X-ray photon produces on average about 1000 electrons in the liquid.

Therefore, the following procedure might be applied to switch to a single electron counting mode. First, with the standard drift field direction the X-ray intensity is decreased down to a single X-ray photon counting level. According to Poisson statistics, this is determined by the fact that the counting efficiency is below 0.2. Then, the direction of the drift field is reversed. This results in the suppression of the primary ionization signal and only the photoemission signal from the first GEM, induced by primary scintillations in the liquid Ar, is detected (see figure 1). Here the first (copper) electrode of the first GEM acts as a photocathode, providing photoelectrons for collection and multiplication in the GEM holes [16]. If the average number of photoelectrons produced by X-ray photon scintillation is much below unity, the detector is operated in a single electron counting mode. This can be further explained as follows.

Taking into account the scintillation yield in liquid Ar (40 photons per 1 keV of deposited energy [17]), typical X-ray photon energy (25 keV), typical quantum efficiency of a metal electrode in the emission region of argon (of the order of $10^{-3}$ [18]), solid angle factor (0.3) and photoelectron backscattering factor (0.3), the number of photoelectrons produced at the first electrode of the first GEM per X-ray photon is roughly estimated to be

$$n_{PE} = 40 \cdot 25 \cdot 10^{-3} \cdot 0.3 \cdot 0.3 = 9 \times 10^{-2} \quad (2).$$

This is not far from the experimental value determined as the ratio of counting efficiencies for the drift field at the reversed and standard directions: $n_{PE}=5\times10^{-2}$. Such a low value of photoemission signal guarantees that the detector is operated in a single electron counting mode with the reversed field, whenever with the standard field direction it is operated in a single X-ray photon counting mode. Accordingly, in this section all the measurements were carried out with the reversed drift field, at a drift field value of $E_{LAr}$=-0.4 kV/cm.

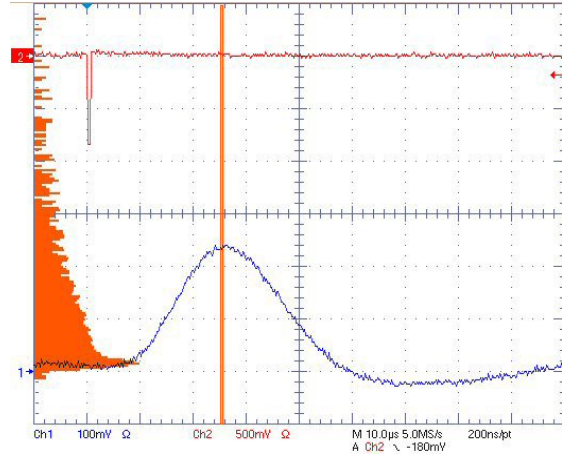

Figure 4. A typical anode signal from the triple-GEM (3GEM) operated in a single electron counting mode in two-phase Ar at 84 K and 0.70 atm at a gain of $4\times10^4$ (bottom). A trigger signal from the X-ray generator (top) and pulse-height spectrum (left) are also shown. The shaping time of the anode signal is 10 μs. The drift field is reversed.

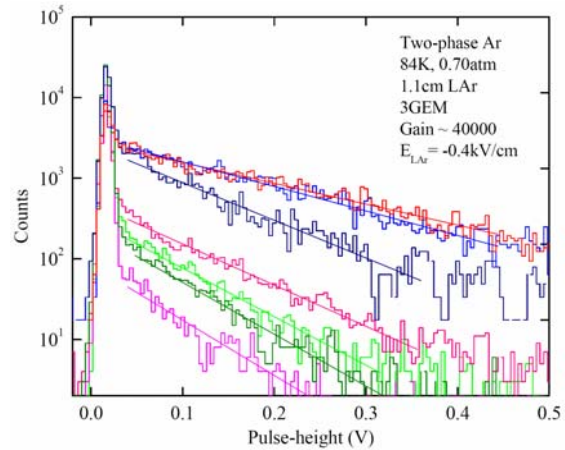

Figure 5. Pulse-height spectra from the triple-GEM (3GEM) as the X-ray source intensity is decreased, in two-phase Ar at 84 K and 0.70 atm at a gain of $4\times10^4$. The drift field is reversed.

Figure 4 illustrates the measurement procedure for single electron counting mode in the two-phase Ar avalanche detector: the pulse-height of the anode signal from the triple-GEM, triggered by the X-ray generator signal, is measured. Figure 5 shows pulse-height spectra measured in this way from the triple-GEM in two-phase Ar at a gain of about $4\times10^4$, as the X-ray source intensity is decreased. The spectra are exponential, which is typical for gas avalanche detectors when counting a few electrons [8]. The curve slope characterizes the average charge recorded, $<q>$. In a single electron counting mode this is simply equal to the gain.

The counting efficiency can be determined as the number of counts under the curve outside the pedestal normalized to the total histogram counts. Figure 6 shows the curve slope as a function of the counting efficiency derived from the spectra of figure 5. One can see that when decreasing the X-ray intensity, the curve slope first decreases and then reaches a plateau at lower efficiencies. This is a signature of operation in a single electron counting mode. Another indication of this is that the average charge on the plateau derived from the curve slope, $<q>$, is consistent with the gain value measured with the standard drift field direction (in section 3).

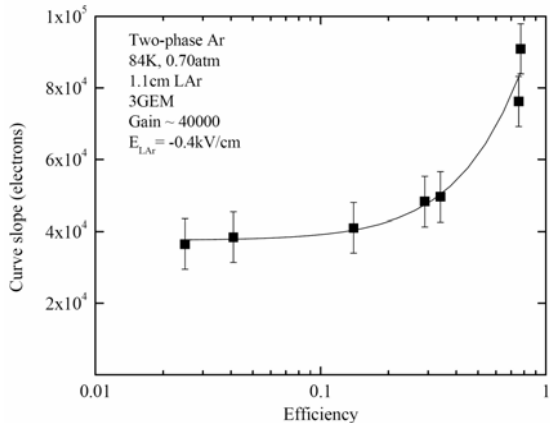

Figure 6. Curve slope of the pulse-height spectrum from the triple-GEM (3GEM), as a function of the efficiency as the X-ray source intensity is decreased, in two-phase Ar at 84 K and 0.70 atm at a gain of $4\times10^4$. The drift field is reversed. The curve slope expressed in electrons is just the average charge detected.

Thus, one may conclude from figure 6 that the detector operates in a single electron counting mode when the counting efficiency is below 0.2. This condition is equivalent to the statement that the average number of photoelectrons per pulse $N_E$ is below 0.2, since according to Poisson statistics the counting efficiency $\varepsilon$ at small $N_E$ is

$$\varepsilon = 1 - \exp(-N_E) \approx N_E \qquad (3).$$

Figure 7 illustrates operation of the two-phase Ar avalanche detector in single electron counting mode at different gains, of $6\times10^3$ and $1.7\times10^4$. The average charges derived from the curve slopes correspond to these gains with an accuracy of 40%, which is within the gain and curve slope measurement accuracies (see figure 2 and 6), again confirming that the detector counts single electrons. At larger gains some minor secondary effects are seen, revealing themselves as a deviation from the exponential function at higher amplitudes, similarly to that observed for the triple-GEM operation in Ar at room temperatures [8].

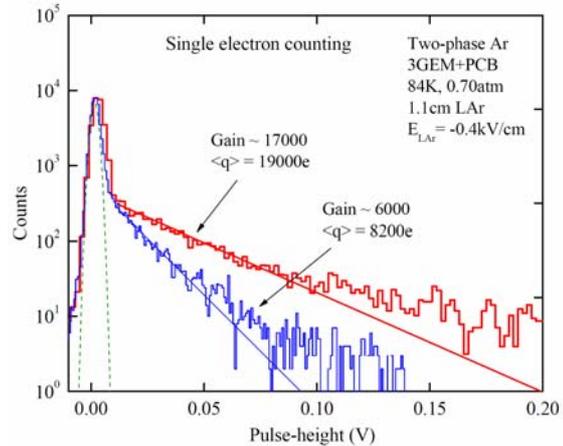

Figure 7. Pulse-height spectra from the triple-GEM (3GEM+PCB) operated in a single electron counting mode at different gains, of $6\times10^3$ and $1.7\times10^4$, in two-phase Ar at 84 K and 0.70 atm. The average charge $<q>$ derived from the curve slope is indicated. A Gaussian fit of the electronic noise spectrum (dashed line) is also shown. The drift field is reversed.

The good single electron signal separation from the pedestal should be emphasized, already at a gain of $6\times10^3$ (figure 7). Here the pedestal width characterizes the electronic noise, corresponding to the Equivalent Noise Charge of $\sigma=940$ e. This value was obtained by fitting the noise spectrum with a Gaussian function (figure 8). The noise spectra were obtained by suppressing the X-ray radiation with a tungsten screen.

One can see from figure 8 that the noise level does not depend on the triple-GEM configuration (compare 3GEM to 3GEM+PCB) and on the gain (compare $1.7 \times 10^4$ to $6 \times 10^3$). This supports the assumption that the electronic noise is basically generated by the relatively long external wires between the triple-GEM electrodes and the preamplifier. This means that in the improved detector the wires between the electrodes and readout electronics should be as short as possible and the noise charge as low as a few hundred electrons may be expected.

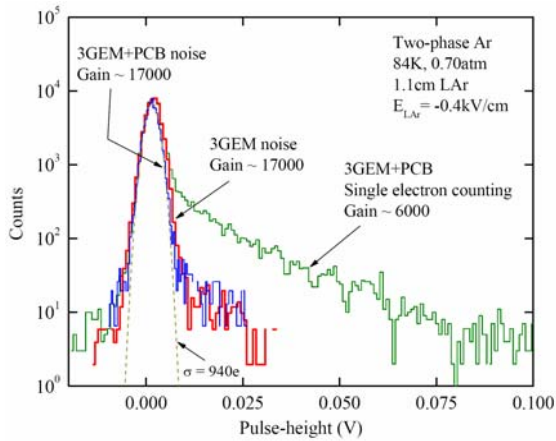

Figure 8. Noise spectra from the triple-GEM in the 3GEM and 3GEM+PCB configurations at a gain of $1.7 \times 10^4$, in two-phase Ar at 84 K and 0.70 atm. For comparison, a single-electron spectrum from the triple-GEM (3GEM+PCB) at a gain of $6 \times 10^3$ is shown. A Gaussian fit to the noise spectrum (dashed line) is also shown. The drift field is reversed.

Analyzing figure 7 and 8, one can estimate the single electron detection efficiency at a certain discrimination threshold when the detector is operated in a self-triggered mode. The normalized exponential distribution describing the single electron spectrum is

$P(q) = \exp(-q/<q>)/<q>$ (4).

Then, the detection efficiency for the charge discrimination threshold $q0$ would be

$$\varepsilon_D = \frac{1}{<q>} \int_{q0}^{\infty} \exp(-q/<q>)\,dq =$$
$$= \exp(-q0/<q>)$$ (5).

For example, for a discrimination threshold of 25 mV corresponding to a charge threshold of $q0=1.4 \times 10^4$ e, the single electron detection efficiency in the 3GEM+PCB configuration at a gain of $1.7 \times 10^4$ would be 50% (see figure 7). Note that this threshold is rather high, corresponding to a signal-to-noise ratio of $15\sigma$, and therefore can be substantially reduced in the improved detector.

It should be remarked that in the present work the noise rate (without the X-ray source) was not minimized: it was of the order of several Hz at the discrimination threshold of 25 mV and gain of several thousands. A detailed study of the noise characteristics of the two-phase Ar avalanche detector will be presented elsewhere [19].

As one can see from figure 6, the counting efficiencies higher than 0.2 correspond to operation in a few electrons counting mode. In particular, the counting efficiency of $\varepsilon=0.76$ corresponds to counting of $N_E=1.4$ electrons on average, according to expression (3). Figure 9 compares the single electron spectrum to that of 1.4 electrons at a gain of $4 \times 10^4$. One can see that due to the different slopes the spectra are well distinguished. The capability of two-phase avalanche detectors to distinguish a single electron spectrum from that of a few electrons is very useful, in particular for characterization of the neutrino spectrum from a nuclear reactor in coherent neutrino scattering experiments [3]. In addition, it might be fundamental for measuring the carrier charge of the electron bubble in cryogenic liquids [20].

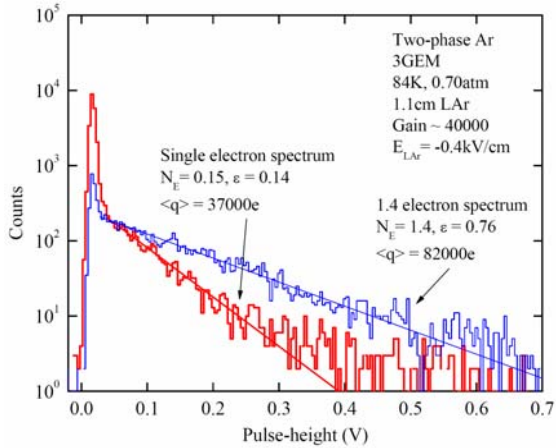

Figure 9. Single electron spectrum and that of 1.4 electrons from the triple-GEM (3GEM) at a gain of $4\times10^4$, in two-phase Ar at 84 K and 0.70 atm. The average number of photoelectrons per pulse ($N_E$) derived from the counting efficiency ($\varepsilon$) and the average charge ($<q>$) derived from the curve slope are indicated. The drift field is reversed.

## 5. Conclusions

The performance of a two-phase Ar avalanche detector was studied. The detector had a 1 cm thick liquid Ar layer and a triple-GEM multiplier operated in saturated vapour above the liquid phase. Triple-GEM gains as high as 20000-40000 were obtained in two-phase Ar and stable detector operation in an avalanche mode for several hours was confirmed.

Successful operation of a triple-GEM multiplier in two-phase Ar, in a single electron counting mode, has also been here demonstrated for the first time. In the gain range of 6000-40000, single electron signals were found to be well separated from noise in the pulse height spectra. The pulse-height resolution of the detector was good enough to distinguish single electron spectra from those of a few electrons.

The successful performance of the two-phase Ar avalanche detector in a single electron counting mode allows one to decrease the energy threshold of the detector down to a few tens of eV, which is fundamental for developing low-background detectors sensitive to nuclear recoils, such as those for coherent neutrino-nucleus scattering and dark matter search experiments.

In the present work, the single electron counting mode was obtained with the reversed drift field, the single electrons being produced by scintillation-induced photoemission from the first GEM acting as a photocathode. It should be noted that a rather similar structure has been recently suggested in the concept of a so-called two-phase avalanche detector with Two-Phase Photoelectric Gate [21]. In this concept, the electron emission through the liquid-gas interface is eliminated by reversing the drift field and the signal is produced by proportional scintillations in the noble liquid recorded at the first GEM acting as a photocathode. Therefore, the results obtained in the present work are also encouraging for the realization of this concept.

## Acknowledgements

The research described in this publication was made possible in part by an INTAS Grant, award 04-78-6744, and through the European Framework 6 project ILIAS, contract number RII3-CT-2004-506222.